\documentclass[12pt, reqno]{amsart}

\usepackage{latexsym, amsfonts, amsmath, amsthm} 
 
\newtheorem{thm}{Theorem}[section]
\newtheorem{lem}[thm]{Lemma}

\newtheorem{prop}[thm]{Proposition}
\newtheorem{cor}[thm]{Corollary}

\theoremstyle{definition}
\newtheorem{definition}[thm]{Definition}

\theoremstyle{remark}
\newtheorem{remark}[thm]{Remark}

\numberwithin{equation}{section}

\newcommand{\ovl}{\overline} 
 
 \newcommand{\inv}{^{-1}}

\newcommand{\ff}{{\mathfrak f}}
\newcommand{\fg}{{\mathfrak g}} 
          \newcommand{\g}{{\mathfrak g}}
\newcommand{\fh}{{\mathfrak h}}

\newcommand{\fr}{{\mathfrak r}}
              
\newcommand{\fs}{{\mathfrak s}}

\newcommand{\fp}{{\mathfrak p}}
\newcommand{\fu}{{\mathfrak u}}

\newcommand{\fz}{{\mathfrak z}}

\newcommand{\C}{{\mathbb C}}

\newcommand{\R}{{\mathbb R}}
\newcommand{\PP}{{\mathbb P}}

\newcommand{\al}{\alpha}
\newcommand{\be}{\beta}
\newcommand{\ga}{\gamma}
\newcommand{\la}{\lambda}

\newcommand{\cG}{{\mathcal G}}
\newcommand{\cH}{{\mathcal H}}

\newcommand{\AND}{\qquad\mbox{and}\qquad}

\newcommand{\Cdn}[1][n]{(\C^d)^{\ot {#1}}}
\newcommand{\Udn}[1][n]{U((\C^d)^{\ot {#1}})}

\newcommand{\Updn}[1][n]{U({d^{#1})}}
\newcommand{\updn}[1][n]{\fu({d^{#1})}}

\newcommand{\SUdd}{SU(d)\times SU(d)}
\newcommand{\sudd}{\su(d)\oplus\su(d)}

\newcommand{\ot}{\otimes}
\newcommand{\ran}{{\rangle}}

\newcommand{\ket}[1]{|{#1}\ran}

\newcommand{\su}{{\fs\fu}} 
\newcommand{\tr}{\mathrm{tr}}   \newcommand{\rank}{\mathrm{rank}} 
\newcommand{\Si}{\Sigma^{\infty}}
\newcommand{\cU}{{\mathcal U}}
\newcommand{\oF}{{\ovl{F}}}

\newcommand{\Hs}{H^\sharp}  \newcommand{\hs}{\fh^\sharp}
\newcommand{\Hf}{H^\flat}  \newcommand{\hf}{\fh^\flat}
\newcommand{\SUU}{SU(d^2)}
\newcommand{\suu}{\su(d^2)}
 
\newcommand{\Fs}{F^\sharp}
\newcommand{\fzs}{\fz^\sharp}

\begin{document}   

\title 
{Universal quantum gates} 
\author{Jean-Luc Brylinski and Ranee  Brylinski}
\address{Department of Mathematics,
        Penn State University, University Park 16802}
\email{jlb@math.psu.edu and   rkb@math.psu.edu}
\urladdr{www.math.psu.edu/rkb}
\thanks{ JLB was supported in part  by NSF  Grant No. 
DMS-9803593. Part of this work was carried out while
both authors were Professeurs Invit\'es at the CPT
and IML of the Universit\'e de la M\'editerran\'ee
in Marseille, France. They are grateful to the Universit\'e de la
M\'editerran\'ee for its hospitality. Also part of this work was done at the
Mini-Symposium on Quantum Computation at College Station
on May 4-6, 2001, where RKB lectured on it. 
We thank Andreas Klappenecker for organizing this symposium.}
 

\begin{abstract} \hskip 1pc  
In this paper we study universality for  quantum gates acting on qudits.
Qudits are states in a Hilbert space of dimension $d$
where $d$ can be any integer $\ge 2$. We determine which $2$-qudit 
gates  $V$ have the properties
(i)  the collection of all  $1$-qudit gates together with $V$ produces all
$n$-qudit gates up to arbitrary precision, or (ii) the collection of all  
$1$-qudit gates  together with $V$ produces all $n$-qudit gates exactly.
We show that (i) and (ii) are equivalent conditions on $V$, and they hold 
 if and only if $V$ is not a primitive gate. Here we say $V$ is primitive if it
transforms any decomposable tensor into a decomposable tensor.  We 
discuss some applications and also relations with work of other authors.
\end{abstract}
 
\maketitle  
\section{Statements  of   main results}
We  determine which $2$-qudit gates $V$ have the property that  all 
$1$-qudit gates together with $V$  form a universal collection, in either the
approximate sense or the exact sense.  Here $d$ is an arbitrary integer
$\ge 2$. Our results are new for the case of qubits, i.e., $d=2$ (which for
many is the case of primary interest). We treat the case $d>2$ as well
because it is of independent interest and requires no additional work.
 
Since Deutsch \cite{D1}  found a universal gate (on $3$ qubits),
universal gates for qubits have been extensively studied.
We mention  in particular the papers \cite{Ba1}, \cite{Ba2}
\cite{D2}, \cite{DV} and \cite{L} which will be further discussed
in \S\ref{sec:examples}.

First we set up some notations.  A \emph{qudit} is a (normalized)
state in the Hilbert space $\C^d$. An \emph{$n$-qudit} is a state in the
tensor product Hilbert space $H=\Cdn=\C^d\ot\cdots\ot\C^d$.
The   \emph{computational basis} of  $H$ is the  orthonormal basis
given by  the $d^n$ classical $n$-qudits
\begin{equation}
\ket{i_1i_2\cdots i_n}=\ket{i_1}\ot\ket{i_2}\ot\cdots\ot\ket{i_n}
\end{equation}
where $0\leq i_j\leq d-1$. The general state in $H$ is a superposition
\begin{equation}\ket{\psi}=
\sum \psi_{i_1i_2\cdots i_n}\ket{i_1i_2\cdots i_n}
\end{equation}
where $||\psi||^2=\sum |\psi_{i_1i_2\cdots i_n }|^2=1$. We say $\psi$ is
\emph{decomposable} when it can be written as a tensor product
$\ket{x_1\cdots x_n}=\ket{x_1}\ot \ket{x_2}\ot\cdots\ot \ket{x_n}$
of  qudits.

A quantum gate on $n$-qudits is a unitary operator $L:\Cdn\to\Cdn$.
These gates form   the  unitary group  $\Udn=\Updn$. 
A sequence $L_1,\dots, L_k$ of  gates constitutes a quantum circuit on
$n$-qudits. The output of that circuit is the product gate $L_1\cdots L_k$.
In practice,  one wants to build   circuits out of gates  $L_i$ which are
\emph{local} in  that they operate on only a small number of qudits,
typically $1$,  $2$ or   $3$.

We can produce local gates in the following way. A $1$-qudit gate $A$ 
gives rise to $n$ different $n$-qudit gates $A(1),\cdots,A(n)$ obtained 
by making $A$ act on the individual   tensor slots. So
\begin{equation}
A(l)\ket{x_1\cdots x_l\cdots x_n}=\ket{x_1}\ot\cdots \ot \ket{x_{l-1}}
\ot A\ket{x_l}\ot \ket{x_{l+1}}\ot\cdots\ot\ket{x_n} \end{equation}
Similarly, for a $2$-qudit gate $B$, we have $n(n-1)$ different
$n$-qudit gates $B(p,q)$ obtained   by making $B$ act on
pairs of  slots. For $B=\sum S_i\ot T_i$ we have $B(p,q)=\sum S_i(p)T_i(q)$.

A basic problem in quantum computation is to find collections of  gates 
which   are universal in the following sense.
\begin{definition} A  collection of $1$-qudit gates $A_i$ and $2$-qudit
gates $B_j$ is called \emph{universal} if, for each $n\geq 2$, 
every $n$-qudit gate can be approximated with arbitrary accuracy by a 
circuit made up of  the $n$-qudit gates produced by the $A_i$ and $B_j$. 
\end{definition}

We also have the stronger notion, which we call   exact  universality.
\begin{definition} A  collection of $1$-qudit gates $A_i$ and $2$-qudit
gates $B_j$ is called \emph{exactly universal} if, for each $n\geq 2$, 
every $n$-qudit gate can be  obtained exactly by a circuit
made up of  the $n$-qudit gates produced by the $A_i$ and $B_j$. 
\end{definition}

In mathematical terms, universality means that  the $n$-qudit
gates produced  by the $A_i$ and $B_j$ generate a dense subgroup of 
$\Updn$, while exact universality means that  these gates generate the 
full group   $\Updn$.

Note that a finite collection of $1$-qudit and $2$-qudit gates can be 
universal,  but it can never be exactly universal, as the group it generates
is countable,  while  $\Updn$ is uncountable.

We now state our main result. We introduce the following terminology.
A  $2$-qudit gate $V$ is \emph{primitive}
if $V$ maps decomposables to decomposables, 
i.e. if $\ket{x}$ and $\ket{y}$ are qudits
then we can find qudits $\ket{u}$ and $\ket{v}$ such that 
$V\ket{xy}=\ket{uv}$. We say $V$ is \emph{imprimitive} when $V$ is not
primitive. Let $P:\Cdn[2]\to \Cdn[2]$ denote the  $2$-qudit gate such that
$P\ket{xy}=\ket{yx}$.

\begin{thm}\label{thm:tg} Suppose we are given a $2$-qudit gate $V$. 
Then the following   are equivalent:
\begin{itemize}
\item[(i)]  the collection of all $1$-qudit gates $A$ together with $V$ is 
universal
\item[(ii)] the collection of all $1$-qudit gates $A$ together with $V$ is 
exactly universal
\item[(iii)] $V$ is imprimitive
\end{itemize}
\end{thm}

We prove Theorem \ref{thm:tg} in \S\ref{sec:3}-\S\ref{sec:norm}. The
implications  (ii)$\Rightarrow$(i)$\Rightarrow$(iii)  are easy. 
The hard part is showing (iii)$\Rightarrow$(ii).
In \S\ref{sec:variant}  we give a variant of   Theorem \ref{thm:tg}.

In \S\ref{sec:prim} we     characterize primitive gates in the following way.
\begin{thm}\label{thm:prim}
$V$ is primitive if and only if $V=S\ot T$ or $V=(S\ot T)P$ for some $1$-qudit 
gates  $S$ and $T$.   Thus $V$ acts  by
 $V\ket{xy}=S\ket{x}\ot T\ket{y}$ or by $V\ket{xy}=S\ket{y}\ot T\ket{x}$.
\end{thm} 
\begin{cor}\label{cor:prim} 
Almost every $2$-qudit gate is imprimitive. In fact the   imprimitive
gates  form a connected open dense  subset of  $\Updn[2]$.
\end{cor}

For the proofs, we use  Lie group theory, including some representation
theory for    compact groups.  For exact universality, we also use some real
algebraic     geometry  (in proving Lemma \ref{lem:gen}).
Our methods can be used to prove a variety of
results on universality and exact universality. We illustrate this is
in \S\ref{sec:variant}.

We thank Goong Chen and Martin  R\"otteler for useful discussions.
and for asking us questions that led, respectively, to the results in  
\S\ref{sec:variant} and the results on exact universality. 

\section{Examples and relations to  works of other authors}
\label{sec:examples} 
In this section, we give examples of primitive and non-primitive gates.
\begin{prop}\label{prop:diag}  
Suppose a $2$-qudit gate $V$ is diagonal in the computational   
basis with $V\ket{jk}=e^{i\theta_{jk}}\ket{jk}$.   Then $V$ is primitive
iff for all $j,k,p,q$ we have
\begin{equation}\label{eq:diag_prim}
\theta_{jk}+\theta_{pq}\equiv \theta_{jq}+\theta_{pk} \pmod{2\pi}
\end{equation}
\end{prop}
\begin{proof}
We apply $V$ to the decomposable tensor
$\ket{\psi}=(\ket{j}+\ket{p})\ot(\ket{k}+\ket{q})$. 
If $V$ is primitive then the result
$V\ket{\psi}=\al_{jk}\ket{jk}+\al_{jq}\ket{jq}
+\al_{pk}\ket{pk}+\al_{pq}\ket{pq}$
 must be  decomposable, where we put $\al_{jk}=e^{i\theta_{jk}}$. Thus
$\al_{jk}\al_{pq}-\al_{jq}\al_{pk}$ vanishes, which amounts to
(\ref{eq:diag_prim}).
Conversely, if (\ref{eq:diag_prim}) holds, we can solve for
scalars $\be_j$ and $\ga_k$ such that $\al_{jk}=\be_j\ga_k$. Then $V=B\ot C$
where  $B\ket{j}=\be_j\ket{j}$  and $C\ket{j}=\ga_j\ket{j}$.
\end{proof} 
 
For example, if  all $\theta_{jk}$ are zero except that
$\theta_{00}\not\equiv 0\pmod{2\pi}$, then  $V$ is imprimitive.
In the case $d=2$,  (\ref{eq:diag_prim}) reduces to   the condition
$\theta_{00}+\theta_{11}\equiv \theta_{01}+\theta_{10}\pmod{2\pi}$        
found in \cite{L}.

In another direction, consider the generalized CNOT gate $X$ given by
$X\ket{ij}=\ket{i,i\oplus j}$ where $i\oplus j$ denotes addition of integers 
modulo $d$. For $d=2$, $X$ is the standard CNOT gate. Then $X$ is 
imprimitive because $X$ transforms the decomposable tensor 
$(\ket{0}+\ket{1})\ot \ket{0})$ into
the indecomposable tensor $\ket{00}+\ket{11}$.
Therefore   the collection of  all $1$-qudit gates together with $X$  is 
exactly universal. This was already proven when $d=2$ in \cite{Ba2}.
 
Here is another kind of controlled gate. Take some $1$-qudit gate $U$ and 
define a $2$-qudit gate $X_U$ by $X_U\ket{0k}=\ket{0}\ot U\ket{k}$ and,
for $j\neq 0$,
$X_U\ket{jk}=\ket{jk}$.
Then $X_U$ is primitive if and only if  $U$ is a scalar operator, i.e.,
$U\ket{x}=e^{i\theta}\ket{x}$. Indeed, for any  $j\neq 0$ we have
$X_U(\ket{j}\ot\ket{x}+\ket{0}\ot\ket{x})=
\ket{j}\ot\ket{x}+\ket{0}\ot U\ket{x}$. This must be  decomposable
if $X_U$ is primitive.
This can only happen if $U\ket{x}=e^{i\theta}\ket{x}$.
Since  $\ket{x}$ is arbitrary,  we see that  $e^{i\theta}$ is     
independent of  $\ket{x}$. Thus  $U$ is a scalar operator.   This 
construction yields many  non-primitive gates  which have finite order.

Another point of view is to consider a $2$-qudit gate $V$ just by itself.
This is interesting  because almost any   $V$   is universal; we call such       
gates   \emph{IU gates} (individually universal). 
This was proven in  \cite{L} and   (for  $d=2$) in \cite {D2}.  More precisely,   
these authors found finitely many open conditions on gates
(e.g., the closure of the subgroup generated by the gate is a maximal torus 
in $\Updn[2]$)  which automatically imply the gate is IU.
In particular, all their gates have infinite order.
 
By theorem \ref{thm:tg},  IU gates are 
imprimitive. There are many gates which are imprimitive
but not IU: for instance,  imprimitive gates which are
diagonal in  the computational basis.

\section{Proof of Theorem \ref{thm:tg} (outline)}\label{sec:3}
We will end up focusing on $2$-qudits, and so we put $G=\Updn[2]$.
We  define $H$ to be the  subgroup of $G$ generated by   the $2$-qudit 
gates $A(1)$  and $A(2)$ for $A\in U(d)$.  Let $F$ be the
subgroup of $G$  generated by $H$, $V$,  $V(2,1)$. 

(ii)$\Rightarrow$(i): obvious

(i)$\Rightarrow$(iii): Suppose $V$ is primitive. We will show that
universality fails for $n=2$, i.e,  $F$  is not dense in $G$. Clearly  $F$ lies in 
the set $L$ of primitive gates. But (a) $L$ is a closed subgroup of $G$ and    
(b) $L\neq G$. Indeed (a) follows easily from the definition of  primitive 
since the decomposable tensors in $\Cdn[2]$ form a closed subset. Also (b)
is true because we already exhibited in \S\ref{sec:examples} some 
$2$-qudit gates which are imprimitive. So   $L$, and hence   $F$,  is not
dense in $G$.

(iii)$\Rightarrow$(ii) takes more work. Here is an outline.
The details   are given in \S\ref{sec:exact} (first step),
\S\ref{sec:dv} (second step),
 \S\ref{sec:lie} (fourth step) and  \S\ref{sec:norm} (fifth step).

\noindent\underbar{First step: }    We give a general abstract result,
Lemma \ref{lem:gen},
which says that if $k$  closed connected subgroups of a compact
group $\cG$ generate a dense subgroup of $\cG$, they must in fact 
generate  $\cG$.  

\noindent\underbar{Second step: }  Using Lemma 4.1 we  reduce  the     
problem to   $n=2$.
 
\noindent\underbar{Third step: }   
$H$ is the set of $2$-qudit gates of the form $S\ot T$. So  $H$ is a closed 
connected Lie subgroup of $G$.   Lemma \ref{lem:gen} suggests that we 
look for    a  closed connected subgroup $H'$ of $G$ such that
\begin{equation}\label{eq:HH} 
\mbox{$H$ and $H'$ generate a dense subgroup  of $G$}  
\end{equation}
The trick is to find  a nice
way to choose  $H'$.  We introduce   the subgroup $H'=VHV\inv$; this is
clearly closed and connected. 
The   next two steps of the proof are devoted to showing 
our group $H'$ satisfies (\ref{eq:HH}).

\noindent\underbar{Fourth step: }   
We will use the Lie algebras $\fg=Lie~G$, $\fh=Lie~H$ and $\fh'=Lie~H'$.
Showing  (\ref{eq:HH}) amounts to showing that $\fh$ and $\fh'$ generate  
$\fg$ as a Lie algebra.  
Let $\fz$ be the Lie subalgebra  generated by $\fh$ and $\fh'$. Then   
$\fh\subseteq\fz\subseteq \fg$.
Using some representation theory, we show abstractly in 
Lemma \ref{lem:liealg} that there is no Lie algebra strictly  in between       
$\fh$ and $\fg$.  Thus $\fz=\fh$   or $\fz=\fg$. 

\noindent\underbar{Fifth step: }   We need to rule out   $\fz=\fh$.
 Clearly $\fz=\fh \Leftrightarrow
\fh=\fh' \Leftrightarrow H=H'\Leftrightarrow V~\textrm{normalizes}~ H$.
But we prove in Proposition \ref{normalizerofH}  that the 
normalizer of $H$ is the set of  primitive gates.  
So $V$ cannot normalize $H$.  Thus $\fz\neq\fh$. 

\noindent\underbar{Sixth step: }   
Thus  $\fz=\fg$.  This proves  (\ref{eq:HH}).
Now (\ref{eq:HH}) and Lemma \ref{lem:gen} imply that $H$ and $H'$ 
generate  $G$. So a fortiori, $H$ and $V$ generate $G$.

\begin{remark}\label{rem:outline}  
(i) We actually proved something stronger than exact universality,
namely that $H$ and $V$ generate $G$. 

(ii) To prove (iii)$\Rightarrow$(i) directly, there is no need for $H'$ or
Lemma \ref{lem:gen}.
We can simply work with $F$. The problem is to show that $F$ is dense in
$G$, which amounts to showing that $\ff=\fg$ where $\ff$ is the Lie
algebra of the closure $\oF$ of $F$ in $G$.  Clearly
$\fh\subseteq\ff\subseteq\fg$ Then we use the same two results, 
Lemma \ref{lem:liealg} and Proposition \ref{normalizerofH}, to show,
respectively, that (a) $\ff=\fh$ or  $\ff=\fg$ and 
(b) $\ff=\fh$ does not happen. 
\end{remark}

\section{First step: From universality to exact universality}
\label{sec:exact}
Our bridge from universality to exact universality is
\begin{lem} \label{lem:gen} 
Let $\cG$  be a  compact Lie group. If  $\cH_1,\dots,\cH_k$ are closed
connected subgroups   and they  generate a dense subgroup of $\cG$, then
in fact they   generate $\cG$.
\end{lem}
\begin{proof} 
We can take $k=2$ since the general case easily reduces to this. Consider 
the subset  $\Sigma=\cH_1\cH_2$ of $\cG$ and its $n$-fold products
$\Sigma^n=\Sigma\cdots\Sigma$. Then $\Sigma,\Sigma^2,\dots$
is an increasing sequence of  subsets whose union, call it  $\Si$, is dense in 
$\cG$. We want to show that there exists $m$ such that $\Sigma^m=\cG$.

To begin with, we observe that   $\Sigma^n$ is  compact  and connected.
This follows as    $\Sigma^n$ is the image of the continuous multiplication 
map $\mu$ from the compact connected set
$(\cH_1\times\cH_2)^{\times n}$ into $\cG$.  So   $\Si$  is connected.
So $\cG$ is connected.

In fact we can conclude much   more using $\mu$. For $\cG$ has  an 
additional structure compatible with its Lie group structure: $\cG$ is a
smooth irreducible real algebraic  variety. (In fact, we can faithfully
represent $\cG$  on some $\C^N$     and then $\cG$ is   
an irreducible closed real algebraic subvariety of the  space of matrices of  
size  $N$.) The subgroups $\cH_1$ and $\cH_2$ are  closed  irreducible 
subvarieties; here we use the connectedness of  $\cH_1$ and $\cH_2$.

Clearly $\mu$ is a morphism of  irreducible real algebraic varieties.
It follows  using the Tarski-Seidenberg theorem that $\Sigma^n$ is a
semi-algebraic set in $\cG$ and its ``algebraic closure" $Z_n$ is irreducible.
Here   $Z_n$ is the unique
smallest closed real algebraic subvariety of $\cG$ which contains 
$\Sigma^n$.  So $Z_1,Z_2,\dots$ is an increasing sequence of  
closed irreducible subvarieties  whose union is dense in $\cG$. It follows, 
by  dimension theory in algebraic geometry, that $Z_p=\cG$ for some $p$.
Since $\Sigma^p$  is semi-algebraic,  the fact $Z_p=\cG$ implies that
$\Sigma^p$  contains an open neighborhood  $\mathcal O$
of one of its points $g$.   (This is the payoff for introducing real algebraic
geometry.)  Now it follows that $\Sigma^{2p+1}$ contains an open
neighborhood $\cU$  of the identity.  Indeed, we take 
$\cU=\mathcal Og\inv$ and notice that that  $g\inv$ lies in $\Sigma^{p+1}$.

We next claim that $\Si=\cG$. First, $\Si$ is open  in $\cG$; this follows 
since   $\Sigma^{2p+1+k}$ contains the open neighborhood 
$\Omega_k=\cU\Sigma^k$  of $\Sigma^k$.   
Second, $\Si$ is clearly a subgroup of $\cG$. 
But $\cG$ is connected and so $\cG$ has no  open subgroup  other than 
itself.  So $\Si=\cG$.

The last paragraph shows that
$\cG$ is  the union of the increasing sequence of  open sets $\Omega_k$.
But $\cG$ is compact, and so this forces $\cG=\Omega_q$ for some $q$.
Hence $\cG=\Sigma^{2p+1+q}$. 
\end{proof}

\section{Second step: Reduction to $n=2$}\label{sec:dv}
 \begin{thm}\label{thm:2eu}
The set of all $2$-qudit gates is exactly universal.
\end{thm}
\begin{proof}  We will apply Lemma \ref{lem:gen} to the $\binom{n}{2}$ 
subgroups  $H(p,q)=\{B(p,q)\,|\, B\in G\}$ of $\Updn$, indexed by pairs 
$(p,q)$  with $p<q$.  Each  $H(p,q)$ is  a connected closed subgroup of 
$\Updn$.  We need to show  that the $H(p,q)$ generate a dense subgroup of
$\Updn$; this amounts to showing that the  Lie algebras of the   $H(p,q)$
generate the  Lie algebra of $\Updn$.  This was done   by DiVincenzo in
\cite{DV} . Although  DiVincenzo only worked in the case $d=2$, his
method  easily extends to the case $d>2$. Thus
 Lemma \ref{lem:gen} applies and tells us the $H(p,q)$  generate $\Updn$.  
\end{proof}
For $d=2$,  Theorem \ref{thm:2eu} was already known by rather  different
methods. It was explained in \cite{Ba2} how to explicitly build any
$n$-qudit gate out of the $n$-qudit gates produced by the 
$1$-qubit gates $A$ together with the  CNOT gate.

\section{Fourth Step:  Analyzing the Lie algebra $\fg$.} \label{sec:lie} 
\begin{lem} \label{lem:liealg} There are no Lie algebras strictly  in between
 $\fh$ and $\fg$.
\end{lem}
\begin{proof} 
We will write elements of $G=\Updn[2]$ and $\fg=\updn[2]$ as matrices of 
size $d^2$,  by using the computational basis of $\Cdn[2]$.  Now $H$ is the 
subgroup of  $G$ of matrices of the form
$h_{S,T}$ where $h_{S,T}=S\ot T$ is the Kronecker product of
unitary      matrices  $S$ and $T$ of size $d$.  

The main idea now is to study $\fg$ as a representation $\pi$ of $K=\SUdd$ 
where $(S,T)$ acts on $\g$ by   $\pi^{S,T}(\xi)=h_{S,T}\,\xi\,h_{S,T}\inv$.
This is useful because if $\fr$ is a Lie subalgebra of $\fg$ and $\fr$ 
contains $\fh$, then the operators $\pi^{S,T}$  preserve  $\fr$. 
So   $\fh\subseteq \fr\subseteq \fg$ as   representations of $K$.
We will show that  there is no representation   of $K$ strictly in
between   $\fh$ and $\fg$.

Now $\fg$ decomposes into a direct sum of irreducible representations of 
$K$. This follows formally since $K$ is a compact Lie group. But also we
can write down the decomposition explicitly. 

To do this, we   observe that each element of  
$\fg$ is a finite sum of   Kronecker products $X\ot Y$ where $X$ lies in 
$\fu(d)$
and  $Y$ lies in $i\fu(d)$. Here $\fu(d)=Lie~U(d)$ is the space of 
skew-hermitian matrices of size $d$. Moreover
$\pi^{S,T}(X\ot Y)=(SXS\inv)\ot(TYT\inv)$.
Thus  $\fg$ identifies naturally with the tensor product
$\fu(d)\ot(i\fu(d))$, where Kronecker product corresponds to tensor 
product. The representation $\pi$ then corresponds to the obvious tensor
product  representation of $K$ on  $\fu(d)\ot(i\fu(d))$. 
As a representation of $U(d)$ under conjugation, $\fu(d)$ decomposes
into  the direct sum of two irreducible  representations: 
$\fu(d)=i\R\,I \oplus\su(d)$, where $I$ is the identity matrix
and $\fs\fu(d)=Lie~SU(d)$ is the space of skew-hermitian matrices of 
trace $0$. Thus we obtain the decomposition
\begin{equation}\label{eq:fg=}
\fg=(i\R\,I \oplus\su(d))\ot (\R\, I\oplus i\,\su(d))=
\fp_0\oplus\fp_1\oplus\fp_2\oplus\fp_3
\end{equation} 
into  four irreducible representations of $K$, where
$\fp_0= i\R\,  I\ot I$,  $\fp_1=\su(d)\ot I$,  $\fp_2=I\ot \su(d)$, and 
$\fp_3=i \,\su(d)\ot\su(d)$.
 
We recognize  $\fh=\fp_0\oplus\fp_1\oplus\fp_2$; this follows since   
$\fh$ consists of matrices of the form 
$X\ot I+I\ot Y$ where $X$ and $Y$ lie  in  $\fu(d)$.
Thus $\fg=\fh\oplus\fp_3$ and so
there is no  representation  of $K$ strictly  in between $\fh$ and $\fg$.
\end{proof}

\section{Fifth Step: The normalizer of $H$}\label{sec:norm}   
We can now show
\begin{prop} \label{normalizerofH} The normalizer of $H$ in $G$ is the group
$L$ of primitive gates.
\end{prop}
\begin{proof} 
We showed in \S\ref{sec:3}  in proving (i)$\Rightarrow$(iii) that
$L$ is a closed subgroup of $G$ with $L$ lying strictly in between $H$ and
$G$.  It follows by Lemma \ref{lem:liealg} that the Lie algebra of $L$ is 
$\fh$. Now,   since  $H$ is a connected Lie group, it follows that
$L$ normalizes $H$.   

For the converse, we
return to our setup in the proof of Lemma \ref{lem:liealg}. 
Let us write $X(1)=X\ot I$ and $Y(2)=I\ot Y$ for any matrices $X$ and $Y$    
of size  $d$.  We identified  $\fh$ as the set of  matrices  
$X(1)+ Y(2)$  of size $d^2$ where $X$ and $Y$ lie  in  $\fu(d)$.

Suppose $R\in G$  normalizes $H$. Then the conjugation action of $R$ on 
$\fg$ preserves $\fh$.   So given any $X,Y\in\fu(d)$, we  have  
\begin{equation}\label{X+Y}
R(X(1)+Y(2))R^{-1}=X'(1)+Y'(2)
\end{equation}
for some $X',Y'\in\fu(d)$.  Then $\tr~X+\tr~Y=\tr~X'+\tr~Y'$ where  
$\tr~X$ is the trace of $X$.  Consequently we can make $X'$ and $Y'$ unique     
by requiring $\tr~X=\tr~X'$ and $\tr~Y=\tr~Y'$. In particular, if 
$X,Y\in\su(d)$, then  $X',Y'\in\su(d)$. In this way, $R$ defines a linear
endomorphism $\ga_R$ of $\sudd$ where $\ga_R(X,Y)=(X',Y')$. 
Clearly $\ga_R$ is  invertible. Moreover
$\ga_R$ preserves the Lie algebra bracket -- this follows using
$[X(1)+Y(2),U(1)+V(2)]=[X,U](1)+[Y,V](2)$.  
Thus $\ga_R$ is a Lie algebra automorphism.

Any Lie algebra  automorphism of $\sudd$ either preserves the two 
summands    or permutes them. This is forced because 
$\su(d)$ is a   simple Lie algebra.  So we have two cases:
either $\ga_R$ preserves the summands so that
$\ga_R(X,0)=(X',0)$ and $\ga_R(0,Y)=(0,Y')$, or $\ga_R$ permutes the
summands so that  $\ga_R(X,0)=(0,Y')$ and $\ga_R(0,Y)=(X',0)$.  
In the latter case, notice that 
$RP$ normalizes $H$ (since  $P$ normalizes $H$) and
$\ga_{RP}=\ga_R\ga_P$ preserves the summands 
(since $\ga_P$ permutes them).  
So either $\ga_R$ or $\ga_{RP}$ preserves the summands.
It is enough to show that $R$ or $RP$ is primitive, since  
$P$ itself is primitive. So we will assume that $\ga_R$ preserves the
summands. Then
\begin{equation}\label{eq:gag} 
RX(1)R\inv=X'(1) \AND RY(2)R\inv=Y'(2)
\end{equation}
                                                                                                                                                                                       
We want to show (\ref{eq:gag}) implies that   $R$ is primitive.
Suppose we have a decomposable $2$-qudit $\ket{xy}$.  We want to show
$R\ket{xy}$ is also decomposable. To do this, we introduce matrices $X$ 
and $Y$ in $\fu(d)$ as follows: $X=ip_x$ and $Y=ip_y$ where $p_x$ is the 
matrix which orthogonally projects $\C^d$ onto the line $\C x$.
Now (\ref{eq:gag}) produces two matrices $X'$ and $Y'$ in $\fu(d)$. 
(Clearly  (\ref{eq:gag}) extends automatically to the case where
$X,Y,X',Y'$ lie in $\fu(d)$, since $\fu(d)=i\,\R I\oplus\su(d)$.)

We claim that $X'$ and $Y'$ are also of the form 
$X'=ip_{x'}$ and $Y'=ip_{y'}$ for some qudits  $\ket{x'}$ and $\ket{y'}$.
This is true   for $X'$
because  $X'$ is skew-hermitian,  $\tr~X'=i$ and  $\rank~X'=1$.  
We  computed the rank of $X'$ in the following way:
(\ref{eq:gag}) implies   $X(1)$ and
$X'(1)$ have the same rank. But $\rank~X(1)=d(\rank~X)=d$ and
$\rank~X'(1)=d(\rank~X')$. Then
\begin{equation}\label{eq:RXYR} 
RX(1)Y(2)R\inv=RX(1)R\inv RY(2)R\inv=X'(1)Y'(2)=-p_{x'}(1)p_{y'}(2)
\end{equation} 
 
Let us apply both sides  of (\ref{eq:RXYR})  to  $R\ket{xy}$. 
The left hand side   gives  $R\ket{xy}$. The right hand side  must be of 
the form  $e^{i\theta}\ket{x'y'}$.  So  $R\ket{xy}=e^{i\theta}\ket{x'y'}$       
is decomposable.
\end{proof}

\section{Proof of Theorem \ref{thm:prim}}\label{sec:prim}
In this section, we   use only the work from \S\ref{sec:lie}-\ref{sec:norm}.
The following result  combined with   Proposition \ref{normalizerofH}   
gives Theorem  \ref{thm:prim}.
\begin{prop} \label{prop:normalizerofH_2}  The normalizer of $H$ in $G$ is  
the union of  $H$ and   $HP$.
\end{prop}
\begin{proof} 
We return to the last phase in the  proof of  Proposition
\ref{normalizerofH}. We showed not only  $R\ket{xy}=e^{i\theta}\ket{x'y'}$
(where $\theta$ depends  on $x,y,x',y'$) but also $x'$ depends only on $x$
while $y'$ depends only on $y$. Furthermore $x$ and $y$ determined $x'$
and $y'$ uniquely up to   phase factors.

We now construct a $1$-qudit gate  $S$ as follows:
we fix  choices of $y$ and $y'$ and then define $S$ by  
$R\ket{xy}=S\ket{x}\ot\ket{y'}$. If we change  our choices of $y$ and $y'$,
then this changes $S$ only by an overall phase factor.
Similarly, we construct a $1$-qudit gate  $T$  by  
$R\ket{xy}=\ket{x'}\ot T\ket{y}$ where this time we fixed choices of $x$ 
and $x'$. Now, for each $\ket{xy}$,  $R\ket{xy}$ coincides with  
$S\ket{x}\ot T\ket{y}$  up to a phase factor which depends on $\ket{xy}$.
It is easy to see that these phase factors are in fact all the same.
Thus $R=e^{i\theta}S(1)T(2)$. So $R$ belongs to $H$.  

This finishes the case where (\ref{eq:gag}) holds. In the other case, where
$RX(1)R\inv=Y'(2)$ and $RY(2)R\inv=X'(1)$, we conclude that $RP$ lies in $H$.
Thus every $R$ normalizing $H$ belongs to either $H$ or $HP$.
The converse is clear.
\end{proof}
\noindent We note that $HP=PH$ since $P$ normalizes $H$.

Using Theorem \ref{thm:prim} we can derive  explicit equations 
characterizing  primitive gates. Let $V_{ij,kl}$ be the matrix coefficients of
$V$ in the computational basis. \begin{cor}  Let $V$ be a $2$-qudit gate.
Then $V$ is primitive  if and only if $V$ satisfies one of the following two 
conditions:
\begin{itemize}
\item[(i)]  $V_{ij,kl}V_{\bar i\bar j,\bar k\bar l}=
V_{i\bar j,k\bar l}V_{\bar ij,\bar k l}$
\item[(ii)] $V_{ij,kl}V_{\bar i\bar j,\bar k\bar l}=
V_{i\bar j,\bar k l}V_{\bar ij,k\bar l}$
\end{itemize}
\end{cor}
\begin{proof}  We will show that   $V$ belongs to $H$ if and only if   
(i) holds, while $V$ belongs to $HP$ if and only if (ii) holds.

We can view $V$ as an element of  $M(d)\ot M(d)$ where $M(d)$ is the 
space of matrices of size $d$.  Now $V$ is \emph{decomposable} 
in this setting if and only if we can find $A$ and $B$ in $M(d)$ such that    
$V=A\ot B$ (so that $V\ket{xy}=A\ket{x}\ot B\ket{y}$). 
Now we recognize (i) as the classical set of quadratic equations which 
characterize when $V$ is decomposable. The point is that $V$ is 
decomposable   only if $V$ belongs to $H$ (the converse is obvious). 
Indeed, if $V=A\ot B$  then, since $V$ is unitary, it follows easily that 
$A=\la S$ and $B=\la\inv T$ where
$\la$ is a positive number and $S$ and $T$ are unitary. So $V=S\otimes T$.
 
On the other hand,  $V$ belongs to $HP$ if and only if  $VP$ belongs to $H$.
But (i) holds for $VP$  if and only if  (ii) holds for $V$.
\end{proof}
\begin{remark} We have a different (and more direct) way of proving 
Theorem \ref{thm:prim}  using  some projective complex algebraic     
geometry.  The  starting point  is to observe that a primitive gate  $V$
induces a holomorphic automorphism of  $\C\PP^{d-1}\times\C\PP^{d-1}$.
\end{remark}

Finally, we prove Corollary \ref{cor:prim}.  The set of imprimitive gates
is $G\setminus L$. This is open in $G$ since we proved $L$ is closed.
The rest requires using results on the  topology of smooth manifolds.
Since $L$ is a  closed submanifold of $G$ with $L\neq G$, it follows that
$G\setminus L$ is dense in $G$. Now connectedness of $G\setminus L$ 
follows   as soon as we check that  $L$ has codimension at least two in $G$.
This is the case because $\dim G=d^4$ and $\dim L=\dim H=2d^2-1$ and so 
the codimension is  $d^4-2d^2+1\geq 9$.

\section{A variant of Theorem \ref{thm:tg}}\label{sec:variant}
In this section  we  consider, in response to a question of G. Chen, what
happens   to Theorem \ref{thm:tg} when we require that the $1$-qudit 
gates  $A$ are \emph{special}, i.e., satisfy   
$\det A=1$. We can prove an analog of  (i)$\Leftrightarrow$(iii): given a  
$2$-qudit gate $V$, the following are equivalent:
\begin{itemize}
\item[(i$'$)] The   collection of all special $1$-qudit gates $A$ 
together  with $V$  is universal.
\item[(iii$'$)] $V$ is imprimitive and $\det V$ is not a root of unity.
\end{itemize}
We cannot get exact universality here because the determinants of the
gates generated by  $A(1),A(2),V,V(2,1)$  are constrained to all be powers of       
$\det~V$. But these powers form only a dense subset of  $U(1)$. So a 
certain set of determinants   never appears.

We can get a full analog of Theorem \ref{thm:tg} in  the following  
way: 
\begin{thm}\label{thm:family} Suppose we are given a family $X$
of $2$-qudit gates $Q_\phi$, indexed by angles $\phi$ modulo $2\pi$,
such that $\det Q_\phi=e^{i\phi}$. 
Then the following   are equivalent:
\begin{itemize}
\item[(i)]  the collection of all special $1$-qudit gates $A$ together
 with $X$
is  universal
\item[(ii)] the collection of all special $1$-qudit gates $A$   together
 with $X$ is exactly  universal
\item[(iii)] at least one $Q_\phi$ is imprimitive
\end{itemize}
\end{thm}

\begin{proof} Each part runs parallel to the proof of Theorem \ref{thm:tg}. 
We  define $\Hs$ to be the  subgroup of $\SUU$ generated by   the
gates $A(1)$  and $A(2)$ for $A$ special; then
$\Hs$ is the set  of gates of the form $S\ot T$ where $S$ and $T$ belong to
$SU(d)$. Let $\Fs$ be the subgroup of $U(d^2)$  generated by $\Hs$ and all 
the gates  $Q_\phi$  and $Q_\phi(2,1)$. 

(ii) $\Rightarrow$(i) is obvious. 

(i) $\Rightarrow$(iii): if (iii) fails, then $\Fs$ lies in  
the group of $L$ of primitive gates. But $L$ is  not dense in $G$.

(iii)$\Rightarrow$(ii):  
We can take $n=2$ as in the proof of Theorem \ref{thm:tg}.
Pick some $Q_\phi$ which is not primitive, and put
$V=Q_\phi$.   Our aim is to show $\Fs=\Updn[2]$.

We claim that $\Hs$ and $\Hf$ generate $\SUU$, where 
we put  $\Hf=V\Hs V\inv$.
Clearly,  $\Hs$ and $\Hf$ are closed connected subgroups of $\SUU$. 
So, by Lemma \ref{lem:gen},  proving the  claim reduces to  showing
that  $\Hs$ and $\Hf$ generate a dense subgroup of $\SUU$.
This amounts to showing that the Lie algebras 
$\hs=Lie~\Hs$ and $\hf=Lie~\Hf$ generate $\fg$.  

Let $\fzs$ be the Lie algebra generated by $\hs$ and $\hf$; then
$\hs\subseteq\fzs\subseteq\suu$. As in the proof of 
Lemma \ref{lem:liealg}, $\fzs$ must be a representation of $K$. So we 
return to the  decomposition (\ref{eq:fg=}). We recognize that
$\suu=\fp_1\oplus\fp_2\oplus\fp_3$ while $\hs=\fp_1\oplus\fp_2$.
Therefore $\suu=\hs\oplus\fp_3$. We  conclude   
$\fzs=\hs$ or  $\fzs=\suu$. 

We want to rule out  $\fzs=\hs$.  
Clearly $\fzs=\hs\Leftrightarrow
\hs=\hf \Leftrightarrow \Hs=\Hf\Leftrightarrow V~\textrm{normalizes}~ 
\Hs$. But $\Hs$ and $H$ have the same normalizer: this follows since
$H$ is the product of $\Hs$ with the scalar $2$-qudit gates, and also
$\Hs$ is the set of  gates in $H$ with determinant equal to $1$.
So Proposition \ref{normalizerofH} tells us that  that $V$ cannot normalize
$\Hs$. Thus $\fzs\neq\hs$.

This proves our claim that $\Hs$ and $\Hf$ generate $\SUU$. 
Therefore $\Fs$ contains $\SUU$. But also $\Fs$ contains a gate 
of each determinant $e^{i\phi}$. So $\Fs=\Updn[2]$.
\end{proof}

Here is a concrete illustration  which was suggested to us by G. Chen.  We
take $d=2$ and consider the gates (written in the computational basis)
\begin{equation}\label{eq:matrices} 
U_{\theta,\phi}=
 \begin{pmatrix} \cos\theta&  -ie^{i\phi}\sin\theta\\
-ie^{-i\phi}\sin\theta& \cos\theta
\end{pmatrix},\ \ \ 
Q_{\phi}=\begin{pmatrix} 1&0&0&0\\ 0&1&0&0\\
0&0&1&0\\0&0&0&e^{i\phi}
\end{pmatrix}
\end{equation}
\begin{cor}  The collection of gates  $U_{\theta,\phi}$ and $Q_\phi$
(where $\theta$ and $\phi$ run through $\R$)  is exactly universal.
\end{cor}

\begin{proof} 
It is  known that the gates  $U_{\theta,\phi}$ generate $SU(2)$.   We can
also see this directly  using  Lemma \ref{lem:gen}. Indeed, for each value 
of $\phi$, the $U_{\theta,\phi}$ form a  closed connected subgroup 
$S_{\phi}$ of $SU(2)$.   Consider the two subgroups $S_0$ and $S_{\pi/2}$.
It is easy to see that  their Lie  algebras generate $\su(2)$. This means
$S_0$ and $S_{\pi/2}$  generate a dense
subgroup of $SU(2)$. So  by  Lemma \ref{lem:gen}, 
$S_0$ and $S_{\pi/2}$  generate $SU(2)$.

Obviously $\det Q_\phi=e^{i\phi}$, and so we get exact universality from
Theorem \ref{thm:family} as soon as we check that some $Q_\phi$ is
imprimitive. In fact, we saw in \S\ref{sec:examples} that  
$Q_\phi$ is always imprimitive,   except of course if $Q_\phi$ is the 
identity.
\end{proof}

\end{document}